\documentclass[aps,prx,twocolumn,groupedaddress,floatfix,longbibliography]{revtex4-2}

\usepackage{graphicx}
\usepackage{dcolumn}
\usepackage{bm}
\usepackage{amsmath}
\usepackage{xcolor}
\usepackage{braket}
\usepackage[utf8]{inputenc}
\usepackage[T1]{fontenc}
\usepackage{hyperref}
\hypersetup{
breaklinks=true,
    colorlinks=true,
    linkcolor=blue,
    citecolor=blue,
    filecolor=magenta,
    urlcolor=cyan,
}

\usepackage{blkarray}
\usepackage{multirow}

\usepackage{natbib}      

\newcommand{\cc}{{\cal C}}

\newcommand{\bG}{{\bf G}}
\newcommand{\br}{{\bf r}}

\begin{document}

\title{Crystal-field effects in the formation of Wigner-molecule supercrystals in moir\'e
TMD superlattices}
\author{Constantine Yannouleas}
\email{Constantine.Yannouleas@physics.gatech.edu}
\author{Uzi Landman}
\email{Uzi.Landman@physics.gatech.edu}

\affiliation{School of Physics, Georgia Institute of Technology,
             Atlanta, Georgia 30332-0430}

\date{27 May 2024}

\begin{abstract}
For moir\'e bilayer TMD superlattices, full-configuration-interaction (FCI) calculations are presented 
that take into account both the intra-moir\'e-quantum-dot (MQD) charge-carrier Coulombic interactions, 
as well as the crystal-field effect from the surrounding moir\'e pockets (inter-moir\'e-QD interactions).
The effective computational embedding strategy introduced here allows for an FCI methodogy 
that enables the complete interpretation of the counterintuitive experimental observations 
reported recently in the context of moir\'e TMD superlattices at integer fillings $\nu=2$ and 4. 
Two novel states of matter are reported: (i) a genuinely quantum-mechanical supercrystal of {\it sliding\/} 
Wigner molecules (WMs) for unstrained moir\'e TMD materials (when the crystal field is commensurate with the
trilobal symmetry of the confining potential in each embedded MQD)
and (ii) a supercrystal of {\it pinned\/} Wigner molecules when the crystal field is incommensurate with the
trilobal symmetry or straining of the whole material is involved.
The case of $\nu=3$ is an exception, in that even the unstrained case is associated with a supercrystal
of pinned WMs, which is due to the congruence of
intrinsic (that of the WM) and external (that of the confining potential of the MQD) $C_3$ point-group
symmetries. Furthermore, it is shown that the unrestricted Hartree-Fock approach 
fails to describe the supercrystal of sliding WMs in the unstrained
case, providing a qualitative agreement only in the case of a supercrystal of pinned WMs.
\end{abstract}   

\maketitle

{\it Introduction:\/}
Earlier studies have revealed a novel fundamental-physics aspect in artificial twodimensional (2D) 
nanosystems in the regime of strong interelectronic correlations, namely formation of quantum Wigner 
molecules (WMs), originally described theoretically \cite{yann99,grab99,yann00,fili01,yann02.2,mikh02,
harj02,yann03,szaf03,yann04,szaf04,roma06,yann06.3,yann07,yang07,yann07.2,yann07.3,umri07,yang08,
roma09,yann15,yann21,erca21.2,urie21,yann22,yann22.2} in 2D semiconductor (parabolic, elliptic, and
double-well) quantum dors (QDs), as well as in trapped ultracold atoms, and subsequently observed 
experimentally in GaAs QDs \cite{yann06,kall08,kim21,kim23}, Si/SiGe QDs \cite{corr21}, planar 
germanium QDs \cite{scar23} and 1D carbon-nanotubes \cite{peck13}.

Remarkably, most recent work \cite{yann23,yann24,crom23} has extended the WM portfolio to the newly 
emerging and highly regarded (due to the potential for fundamental-physics discoveries and for 
advancing quantum-device applications) field of transition-metal dichalcogenide (TMD) moir\'e 
materials and superlattices. Specifically, TMD moir\'e superlattices are highly pursued for solving 
the question of scalability in quantum computer architectures. 

\begin{figure}[t]
\centering\includegraphics[width=8.0cm]{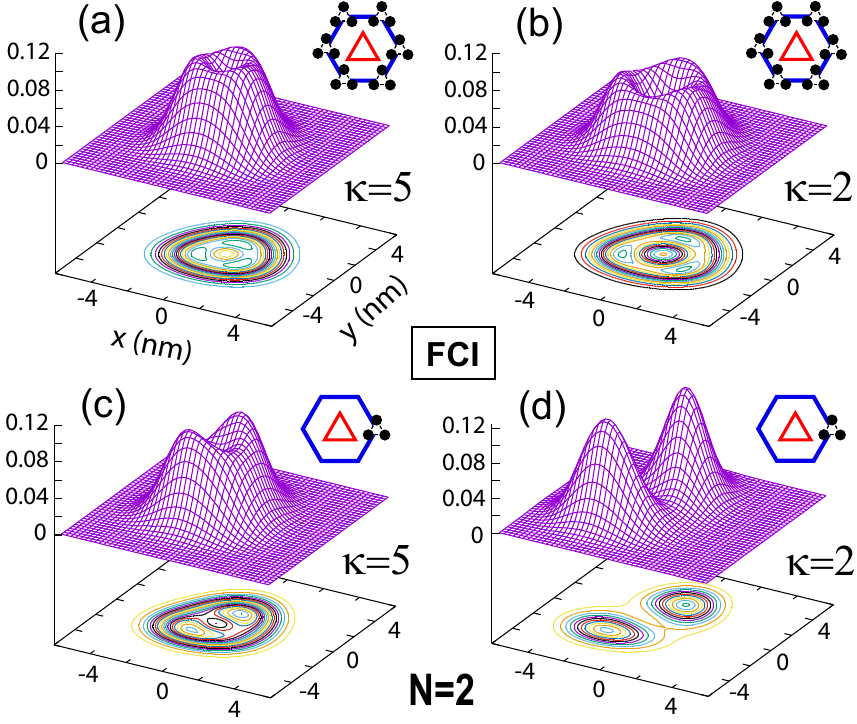}
\caption{
FCI ground-state (with total spin $S=0$) charge densities for $N=2$ holes in the embedded central MQD (red 
triangle in the schematics) taking into consideration the crystal-field effect generated by the charge 
carriers in the surrounding six moir\'e pockets (see blue hexagon, with radius $a_M$, in the 
schematics). 
(a) and (c) $\kappa=5$ in agreement with the case of a TMD moir\'e superlattice in hBN environment.
(b) and (d) were calculated, for purpose of comparison, with $\kappa=2$, corresponding to a stronger 
Coulomb repulsion.
In (a) and (b), all six surrounding MQDs (see schematics) were populated with $Q=Ne$ charge 
carriers. In (c) and (d), only one of the surrounding MQDs (the one to the right, see schematics) has 
been populated with $Q=Ne$ charge carriers. The charge-carrier distributions in each surrounding MQD 
have been represented by three point charges $Q/3$ placed at the apices of an equilateral triangle 
(see schematics), with radius $a_M/6$. 
Remaining parameters: effective mass $m^*=0.90 m_e$, moir\'e lattice constant $a_M=9.8$ nm, 
depth of a moir\'e pocket $v_0=10.3$ eV, and $\phi=20^\circ$.
Charge densities are in units of 1/nm$^2$.
}
\label{chdn2ci}
\end{figure}

Building on the demonstrated formation \cite{yann23} of WMs in the isolated moir\'e pockets [most 
often referred to as moir\'e quantum dots (MQDs)] at integer fillings, we investigate here the effects
arising from embedding such single MQDs in the moir\'e superlattice structure, which is the actual 
system that was addressed experimentally \cite{crom23}. Specifically, this paper adresses the key
question whether the experimental observations in the unstrained-lattice case do
reflect formation of a collective WM supercrystal, i.e., a novel, more complex insulating phase of matter
beyond the well-known generalized Wigner crystal, or the trivial case of a collection of independent
MQDs. The present investigations (see below) support formation of a WM supercrystal \footnote{
We note that the experimental results for the unstrained case in Ref.\ \cite{crom23} were necessarily 
interpreted with formation of sliding WMs in {\it isolated\/} MQDs (for $N=2$ confined charge 
carriers, see Refs.\ \cite{yann23,crom23}; for $N=4$ confined charge carriers, see Ref.\ \cite{yann23}), 
given the fact that the mean-field Hartree-Fock approach (that allows calculations for a large number of 
charge carriers distributed over several moir\'e pockets) is unreliable
\cite{yann24,crom23,fuku81,lowe81,pald85} for 
describing sliding WMs.} having sliding WMs as building blocks for $\nu=2$ or $\nu=4$.
The filling $\nu=3$ is an exception having a pinned WM as a building block.

(We note that we reserve the term "rotating WM" to describe the azimuthal degree of freedom of WMs 
forming only in a  circularly symmetric confinement 
\cite{yann00,yann07,yann03,yann06.3,yann02.2,yann04,yann07.3,yann21,roma06,yann06.2}
(thus exhibiting a good total angular
momentum), whereas the TMD moir\'e pockets are distinguished by an azimuthal $C_3$  trilobal symmetry (see
discussion of the confining potential below), which does not preserve angular momenta.)

Specifically, this paper focuses \footnote{The main text focuses on charge densities (first-order correlations).
Results for conditional probability distributions (second-order correlations) are presented in Appendix \ref{a1}
and Appendix \ref{a7}.}
on the effects on WM formation in a given MQD resulting from the 
crystal-field generated by the charge carriers (of same nature) in the surrounding MQDs (moir\'e 
pockets) located at the apices of a regular hexagon in the TMD moir\'e superlattice. 
This approach is inspired by the well-known crystal-field theory
\cite{beth29,vlec32} in molecular physics, and is applicable here because the moir\'e lattice constant 
$a_M$ is much larger than the extent of the Wigner molecule within a single MQD.   
Two different computational methodologies will be used in this endeavor, namely: 
(i) the full configuration interaction (FCI) \cite{lowd55,shav98,yann03,szaf03,yann06.2,ront06,yann07,
yann07.3,yann09,yann15,yann21,yann22.2,yann22.2,yann22.3,szabo} and
(ii) the spin-and-space unrestricted Hartree-Fock (sS-UHF)
\cite{yann99,yann02,yann02.2,yann07,yann00.2,gold24}.

{\it Many-body Hamiltonian (including crystal field from surrounding moir\'e pockets):\/}
Following earlier established literature \cite{macd18,ange21,fu20,yann23,crom23}, we approximate the 
potential of the 2D TMD moir\'e superlattice that confines the extra charge carriers as 
\begin{align}
V(\br) = -2 v_0 \sum_{i=1}^3 \cos(\bG_i \cdot \br + \phi),
\label{mpot}
\end{align}
where $\bG_i=[ (4\pi/\sqrt{3}a_M) ( \sin(2\pi i/3), \cos(2\pi i/3) ) ]$ are the moir\'{e} reciprocal
lattice vectors. The materials specific parameters of $V(\br)$ are $v_0$ (which can also be
experimentally controlled through voltage biasing), the moir\'{e} lattice constant $a_M$, and the
angle $\phi$. $a_M$ is typically of the order of 10 nm, which is much larger than the lattice
constant of the monolayer TMD material (typically a few \AA). 

The parameter $\phi$ controls the strength of the trilobal $C_3$ anisotropy in
each MQD potential pocket. 
This trilobal anisotropy can be seen by expanding $V(\br)$ in Eq.\ (\ref{mpot}) in powers of $r$, 
and defining an approximate confining potential, $V_{\rm MQD}(\br)$, for a single MQD as follows:
\begin{align}
\begin{split}
V_{\rm MQD}(\br) \equiv V(\br) + 6 v_0 \cos(\phi)  \approx m^* \omega_0^2 r^2/2 +
\cc \sin(3 \theta) r^3.
\end{split}
\label{vexp}
\end{align}
with $m^*\omega_0^2=16 \pi^2 v_0 \cos(\phi)/a_M^2$, and
$\cc=16 \pi^3 v_0 \sin(\phi)/( 3\sqrt{3}a_M^3)$; $m^*$ is the effective mass and the expansion of
$V(\br)$ can be restricted to the terms up to $r^3$. $(r, \theta)$ are the polar coordinates of
the position vector $\br$.

The effective many-body Hamiltonian associated with a given MQD embedded in the moir\'e superlattice 
is given by
\begin{align}
H_{\rm MB} = H_{\rm MQD} + H_{\rm CF},
\label{mbh}
\end{align}
where
\begin{align}
H_{\rm MQD} = \sum_{i=1}^N \left\{ \frac{{\bf p}_i^2}{2 m^*} +V_{\rm MQD}(\br_i) \right\} +
\sum_{i<j}^N \frac{e^2}{\kappa |\br_i-\br_j|},
\label{hmqd}
\end{align}
with $\kappa$ being the dielectric constant, and
\begin{align}
H_{\rm CF} = \sum_{i=1}^N \sum_{j=1}^{3{\cal M}} 
\frac{eQ}{3\kappa |\br_i - {\bf R}_j|} 
\label{vcf}
\end{align}
is the crystal-field potential from the surrounding moir\'e pockets, ${\cal M}$ denoting the
number of surrounding moir\'e pockets populated with charge carriers, and the ${\bf R}_j$'s being the
positions of the point charges that mimick the charge carrier distributions. In each one of the 
surrounding moir\'e pockets, we use three point charges at the apices of an equilateral triangle with 
a total charge $Q=Ne$, thus respecting the $C_3$ trilobal symmetry; see solid dots in the schematics 
in the figures.

A brief outline of the FCI and sS-UHF methodologies, used to solve the corresponding many-body
Schr\"{o}dinger equation, is presented in Appendices \ref{a3}-\ref{a5}, including
Refs.\ \cite{math22,kime93}.

Before proceeding with the computational results, we point out that the arrangement having the 
equilateral three-point-charge distributions in all the six surrounding moir\'e pockets [see schematic 
in Fig.\ \ref{chdn2ci}(a)] displays an overall $C_6$ point-group symmetry, which is commensurate with 
the trilobal $C_3$ symmetry associated with the potential confinement [Eq.\ (\ref{vexp})] of a 
single MQD. Below, we will investigate the effect of both such a commensurate crystal field, as well as 
the effect of an incommensurate crystal field (IcCF) generated by placing the three-point-charge 
distribution in only one of the six surrounding moir\'e pockets [see schematic in Fig.\ 
\ref{chdn2ci}(c)].  

\begin{figure}[t]
\centering\includegraphics[width=8.0cm]{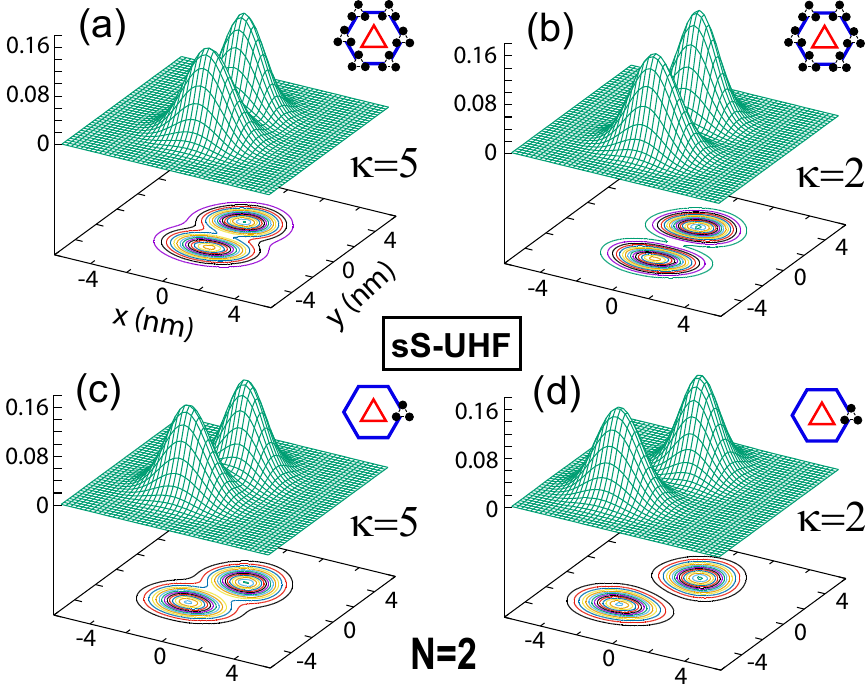}
\caption{
sS-UHF ground-state (with total-spin projection $S_z=0$) charge densities for $N=2$ holes in the central 
MQD (red triangle in the schematics) taking into consideration the crystal-field effect generated by the 
charge carriers in the surrounding six moir\'e pockets (see blue hexagon, with radius $a_M$, in the 
schematics). The information in this figure is organized in the same way as in Fig.\ \ref{chdn2ci}. 
}
\label{chdn2uhf}
\end{figure}

{\it Results for filling factor $\nu=2$:\/}
The FCI and sS-UHF charge densities (CDs) for a moir\'e TMD (e.g., WS$_2$ \cite{crom23}) superlattice at integer 
filling factor $\nu=2$ calculated using the many-body Hamiltonian (\ref{mbh}) are displayed in Fig.\ \ref{chdn2ci} 
and Fig.\ \ref{chdn2uhf}, respectively. The configuration of the equilateral three-point-charge distributions 
that mimick the crystal-field effect from the surrounding six moir\'e pockets is illustrated in the
schematics drawn in the upper right corner of each frame; see figure captions for details and parameters. 
The total charge associated with each equilateral three-point-charge distribution is $Q=2e$. (Note that 
$Q/e$ can be taken different than $\nu$ in the case of a noninteger filling.) 

The FCI calculations [see Figs.\ \ref{chdn2ci}(a) and \ref{chdn2ci}(b)] demonstrate that the full crystal-field
effect (all six surrounding moir\'e pockets are contributing) maintains the charge densities that are 
associated with sliding WMs. Namely, the CDs preserve the $C_3$ symmetry of the confining single-moir\'e-pocket 
potential, as was the case with the isolated MQDs \cite{yann23}. Note that even a stronger Coulomb repulsion 
($\kappa=2$ instead of $\kappa=5$, which corresponds to the experimental setup) does not destroy the 
the $C_3$ symmetry; see Fig.\  \ref{chdn2ci}(b). This symmetry-preserving behavior is 
counterintuitive for a crystalline material in that the CDs necessarily exhibit three modest humps instead of 
the naive expectation of a symmetry-breaking double hump associated with two Wigner-crystal-type 
\cite{wign34,wign38,wang21,yazd24} localized charge carriers (one hump per charge carrier).
Note further that this counterintuitive behavior is in remarkable agreement with the recent experimental 
findings \cite{crom23} for the unstrained case at filling $\nu=2$.

\begin{figure}[t]
\centering\includegraphics[width=8.0cm]{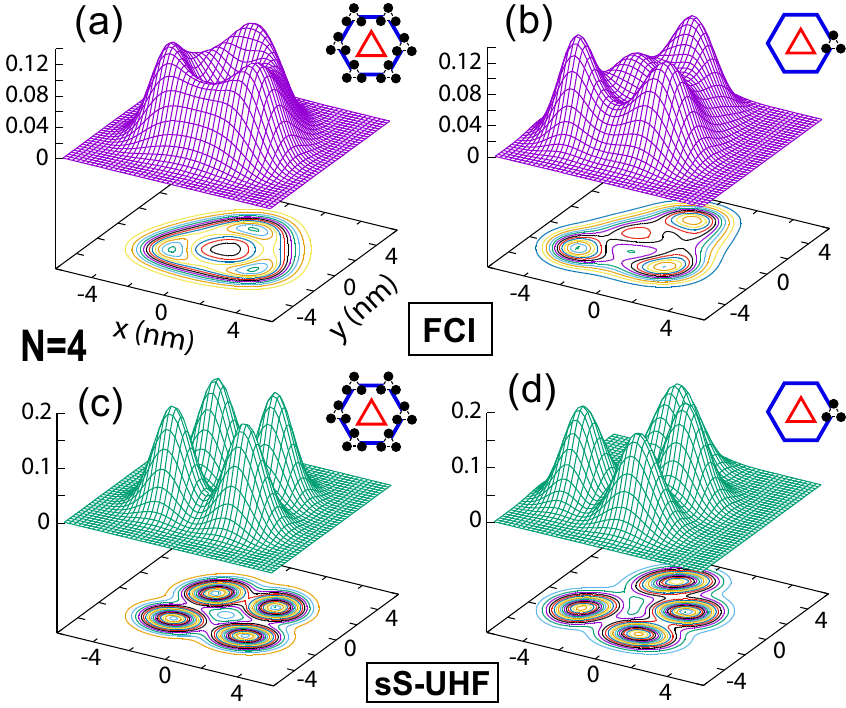}
\caption{
Charge densities for $N=4$ holes in the central MQD (red triangle in the schematics) taking into 
consideration the crystal-field effect generated by the charge carriers in the surrounding six moir\'e 
pockets (see blue hexagon, with radius $a_M$, in the schematics).
(a) and (b) FCI ground-states (with total spin $S=1$ and total-spin projection $S_z=0$). 
(c) and (d) sS-UHF ground-states (with total spin projection $S_z=0$).
 $\kappa=5$ in all panels.
 In (a) and (c), all six surrounding MQDs (see schematics) were populated with $Q=Ne$ charge
carriers. In (b) and (d), only one of the surrounding MQDs (the one to the right, see schematics) has
been populated with $Q=Ne$ charge carriers. 
All other parameters and notations are the same as those in Figs.\ \ref{chdn2ci} and \ref{chdn2uhf}.
}
\label{chdn4}
\end{figure}

The CDs in Figs.\ \ref{chdn2uhf}(a) and \ref{chdn2uhf}(b) demonstrate that the sS-UHF is
unable to properly describe the unstrained TMD moir\'e supercrystal at filling $\nu=2$. Indeed, these
sS-UHF CDs exhibit a pinned WM with two well-localized humps in accordance with the traditional 
Wigner-crystal-type \cite{wign34,wign38,wang21,yazd24} expectation (one hump per charge carrier). 
We stress again that these sS-UHF CDs do break the $C_3$ symmetry of the single-moir\'e-pocket confinement and 
disagree with the experimental observations \cite{crom23}.    

Heretofore, we used a crystal field with $C_6$ symmetry (associated with that of the six surrounding moir\'e
pockets), which is commensurate with the $C_3$ symmetry of the confinement of the embedded MQD. 
A natural question 
arising at this point concerns the effect of a crystal field that has a symmetry non-commensurate with the
$C_3$ symmetry of the single MQD. Experimentally, such a non-commensurability was generated \cite{crom23} by
deforming the moir\'e lattice through straining. Here, we generate a non-commensurability situation by 
maintaining the equilateral three-point-charge configuration only in a single moir\'e pocket [the one on the
right; see schematics in Figs.\ \ref{chdn2ci}(c,d) and \ref{chdn2uhf}(c,d)].       

From the FCI CDs in Figs.\ \ref{chdn2ci}(c,d), it is apparent that the IcCF 
produces an azimuthally {\it pinned\/} WM exhibiting two antipodal humps. These pinned-WM CDs do 
break the $C_3$ symmetry of the single MQD. In addition, the extent of localization for each charge carrier 
becomes more pronounced for stronger Coulomb repulsion; compare the case of $\kappa=5$ [Fig.\ 
\ref{chdn2ci}(c)] to $\kappa=2$ [Fig.\ \ref{chdn2ci}(d)]. Concerning the corresponding sS-UHF CDs, it is
obvious that, in qualitative agreement with the FCI result, the IcCF CDs result also in pinned two-humped 
WMs, exhibiting, nevertheless, a stronger charge-carrier localization; contrast Fig.\ \ref{chdn2ci}(c)
to Fig.\ \ref{chdn2uhf}(c) and Fig.\ \ref{chdn2ci}(d) to Fig.\ \ref{chdn2uhf}(d). We mention here that our
IcCF FCI CDs are in agreement with the recent experimental observations \cite{crom23} of two-humped 
pinned WMs per moir\'e pocket in the case of {\it strained\/} moir\'e superlattices at a filling $\nu=2$.

{\it Results for filling factor $\nu=4$:\/}
The ability of the crystal-field with the commensurate $C_6$ symmetry to preserve, and even enhance, 
the formation of a quantum sliding WM in the central MQD is revealed in an even more spectacular way in
the case of the filling factor $\nu=4$, where Ref.\ \cite{crom23} has reported within each MQD the 
observation of CDs with 3 humps ($C_3$ symmetry) for the unstrained lattice, but with 4 humps
(broken symmetry) for the strained lattice. Our FCI CDs that take into account the crystal-field effect
are in agreement with the experiment; see the 3-hump, sliding-WM CD in Fig.\ \ref{chdn4}(a) (commensurate 
crystal field) and the 4-hump, distorted-pinned-WM CD in Fig.\ \ref{chdn4}(b) (IcCF). We further note 
that the sS-UHF fails to describe the sliding WM, yielding instead a broken-$C_3$-symmetry CD (pinned WM) 
with 4 well defined humps even for the case of the commensurate crystal field; see Fig.\ \ref{chdn4}(c). 
In the case of an IcCF, the sS-UHF agrees qualitatively with the FCI CD as it describes a deformed, 
4-hump, pinned WM; compare Fig.\ \ref{chdn4}(b) and Fig.\ \ref{chdn4}(d), the charge-carrier localization 
being more pronounced in the sS-UHF CD.

\begin{figure}[t]
\centering\includegraphics[width=8.0cm]{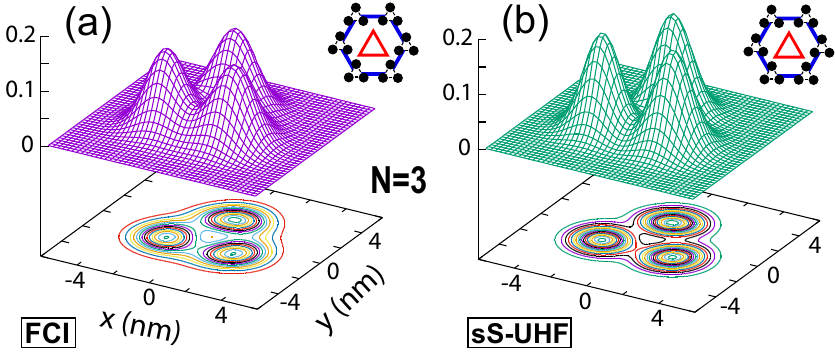}
\caption{
Charge densities for $N=3$ holes in the central MQD (red triangle in the schematics) taking into 
consideration the crystal-field effect generated by the charge carriers in the surrounding six moir\'e 
pockets (see blue hexagon, with radius $a_M$, in the schematics).
(a) FCI ground-states (with total spin $S=1/2$ and total-spin projection $S_z=1/2$). 
(b) sS-UHF ground-states (with total spin projection $S_z=1/2$).
$\kappa=5$ in both panels.  All other details are the same as those in Figs.\ \ref{chdn2ci}-\ref{chdn4}.
}
\label{chdn3} 
\end{figure}

The central role of the overall hexalobular $C_6$ symmetry in enhancing the formation of sliding WMs
is further demonstrated by our calculations for $N=4$ holes using a point-charge arrangement in the surrounding
moir\'e pockets that is antiparallel to that employed in Fig.\ \ref{chdn4}(a). We find that the effect of such a 
flip is minimal, with the central charge density retaining the same shape and orientation as in Fig.\ 
\ref{chdn4}(a); see Fig.\ \ref{S1} in Appendix \ref{a6}.

{\it Results for filling factor $\nu=3$:\/}
Compared to the cases of $\nu=2$ and $\nu=4$, the case of $\nu=3$ is exceptional, a fact that was reported 
already in Ref.\ \cite{yann23} where formation of WMs was investigated for the case of an isolated MQD in 
the absence of a crystal field from the surrounding moir\'e pockets.
Specifically, the 3-hump pinned WM in the central MQD has a $C_3$ symmetry which coincides with that
of the confining potential. This congruence of intrinsic and external point-group symmetries prohibits 
the formation of a sliding WM and yields a strongly pinned WM. The commensurate crystal field does not 
alter this CD distribution, a fact that is reflected in the FCI calculation [see Fig.\ \ref{chdn3}(a)].
In the $\nu=3$ case, the sS-UHF result is in qualitatively agreement with the FCI one [see Fig.\ 
\ref{chdn3}(b)], with the charge-carrier localization, however, being more pronounced in the sS-UHF 
case. Again, it is remarkable that our FCI results (including the crystal-field effect) for $\nu=3$ are 
in excellent agreement with the experiment \cite{crom23}.

\textcolor{black}{
{\it Explicit consideration of strain:\/} Heretofore, we focused on the influence of the symmetry of the
crystal field on the formation of sliding or pinned WMs. Another factor is the presence of strain in the
material. This case is examined in some detail in Appendix \ref{a2}, where Fig.\
\ref{strain} displays the charge densities for $N=2$ holes for values of the strain parameter
$\delta=0.01$, 0.03, 0.1, and 0.15; $\delta$ is defined as in Ref.\ \cite{crom23}.
It is seen that a well defined pinned WM is formed as a function of increasing $\delta$, in close agreement
with the experiment. It is remarkable that the effect of the breaking of the $C_3$ symmetry in the WM charge
density can be resolved for small values of $\delta$, as small as 0.03. 
}

{\it Summary:\/}
Taking into consideration (via a computational embedding scheme)
both the intra-moir\'e-QD charge-carrier interactions in the embedded MQD, 
as well as the crystal-field effect from the surrounding moir\'e pockets (inter-moir\'e-QD interactions), 
and using the full-configuration-interaction methodology, we offered a complete interpretation of the 
counterintuitive experimental observations reported in Ref.\ \cite{crom23} in the context of moir\'e
TMD superlattices at integer fillings $\nu=2$ and 4. In particular, our 
results demonstrate that these experimental observations reflect the interplay between two novel
states of matter; namely, (i) a supercrystal of sliding Wigner molecules \footnote{
When the QD confinement possesses circular symmetry, the sliding WMs were earlier referred to
as rotating WMs or rotating electron molecules \cite{yann07,yann03}.} 
for unstrained moir\'e TMD materials (when the crystal field is commensurate with the trilobal
symmetry of the confining potential in each MQD) and (ii) a supercrystal of pinned Wigner molecules 
when strain is involved and the crystal field is non-commensurate with the trilobal
symmetry of the confining potential in each MQD. The case of $\nu=3$ is an exception, in that both
unstrained and strained cases produce a supercrystal of pinned WMs; this is due to the congruence of 
intrinsic (that of the WM) and external (that of the confining potential of the MQD) $C_3$ point-group 
symmetries.

In summary, our analysis uncovers the physical realization of switching transitions between 
the above-mentioned sliding and pinned Wigner-molecule phases, anticipated earlier 
\cite{yann02.2,yann03,yann07,yann11} 
within the context of fundamental-physics investigations concerning the 
interplay between symmetry breaking and symmetry restored many-body 
theoretical approaches. It is remarkable that such sliding-WM $\Leftarrow\Rightarrow$ pinned-WM  
transitions can be brought about, within a common experimental setup, through readily controlled 
manipulation of a single characteristic material parameter (i.e., strain) of the TMD moir\'e system.
   
Furthermore, we demonstrated that the UHF approach (invoked in Refs.\ \cite{crom23,redd23}) fails 
to describe the genuinely quantum-mechanical supercrystal of sliding WMs in the unstrained 
case, providing a qualitative agreement only in the case of a supercrystal of pinned WMs \footnote{
We remark (following one of the referees' suggestion)  that the comparative methodology that we used here
of employing FCI calculations as benchmarks for assessing the adequacy of UHF calculations, could be 
adopted for assessing advanced multi-configurational (MC)  quantum-chemical computational techniques 
[e.g.,  MC Self-Consistent Field, CASSCF, or the Complete-Active-Space Self-Consistent Field (CASSCF)], 
as well as for improving such techniques (e.g.,  introducing 'dynamic' correlations via perturbative
contributions), thereby extending the range of computationally accessible multi-particle systems}.

This work has been supported by a grant from the Air Force Office of Scientific Research (AFOSR)
under Contract No. FA9550-21-1-0198. Calculations were carried out at the GATECH Center for
Computational Materials Science. We thank the referees of the paper for their constructive comments.

\appendix

\section{CONDITIONAL PROBABILITY DISTRIBUTIONS AND SLIDING WIGNER MOLECULES}
\label{a1}

 The unrestricted Hartree-Fock method employs different space orbitals for the two different spin directions. 
 Thus, the  sS-UHF preserves the spin projection, but allows the total-spin and space symmetries (i.e., the 
 $C_3$ trilobal  symmetry in the potential pockets of a TMD moir\'e superlattice) to be broken, resulting in the
 appearance of a pinned WM; see Figs.\ \ref{chdn2uhf}(a),  \ref{chdn2uhf}(b), and \ref{chdn4}(c) \footnote{An
 analogous discussion in the case of a twodimensional rotational confining potential was presented in 
 Ref.\ \cite{yann07} in the context of semiconductor quantum dots.} 
 The many-body sS-UHF wave function is a single Slater determinant. 
 
The FCI method reproduces the exact many-body solution by providing a many-body wave function that is a linear
superposition of Slater determinants. As shown in this paper, the FCI charge densities preserve the $C_3$ trilobal 
symmetry of the moir\'e potential pocket, even in the presence of the $C_6$ hexalobular TMD crystal field; 
 see Figs.\ \ref{chdn2ci}(a),  \ref{chdn2ci}(b), and \ref{chdn4}(a). 
 
 The $C_3$-symmetry preserving charge densities in  Figs.\ \ref{chdn2ci}(a),  \ref{chdn2ci}(b), and \ref{chdn4}(a)
 do not suggest by themselves any relation to WMs. Nevertheless, for these cases, we used the term "sliding WM" 
 because the intrinsic structure of the associated FCI wave functions consists of a superposition of WM 
 configurations that exhibit quantum fluctuations in the azimuthal orientation (thus the term "sliding"). That this
 is the case is promptly revealed through the inspection of second-order correlations, often referred to as
 spin-resolved conditional probability distributions (SR-CPDs). 
 
\begin{figure}[t]
\centering\includegraphics[width=8.1cm]{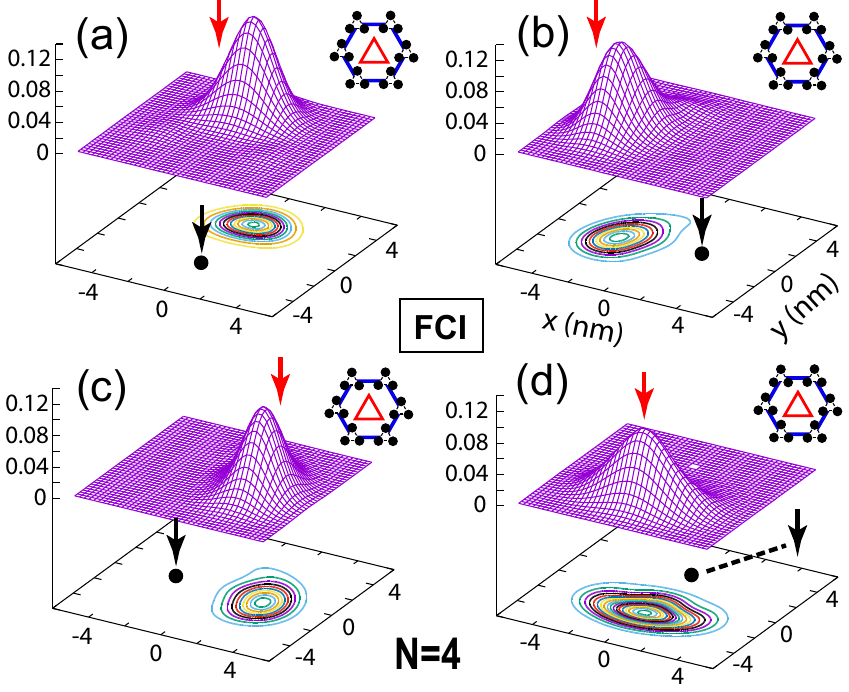}
\caption{
FCI spin-resolved CPDs ($P_{\downarrow\downarrow}$)  for the ground state (with total spin $S=1$ and 
total-spin projection $S_z=0$) for $N=4$ holes in the central MQD taking into 
consideration the crystal-field effect generated by the charge carriers in the surrounding six moir\'e 
pockets (see blue hexagon, with radius $a_M$, in the schematics). $\kappa=5$ in all four panels. 
 The black arrows indicate the chosen down spin of the fixed hole which is situated at the four
 different positions $\br_0$ marked by a solid black dot: (a) $\br_0$=(0,-2.09 nm), (b) $\br_0$=(2.4 nm, 0),
 (c) $\br_0$=(0, 2.61 nm), and $\br_0$=(-2.4 nm, 0). The red arrows indicate that the plotted surfaces reveal 
 the distributions of a second spin-down hole. All other parameters are the same as those in Fig.\ 
 \ref{chdn4}(a), which portrays the corresponding charge density. The CPDs are in units of
 1/nm$^4$ and are normalized to 1/nm$^2$. 
}
\label{cpdn4dd}
\end{figure}

\begin{figure*}[t]
\centering\includegraphics[width=12.0cm]{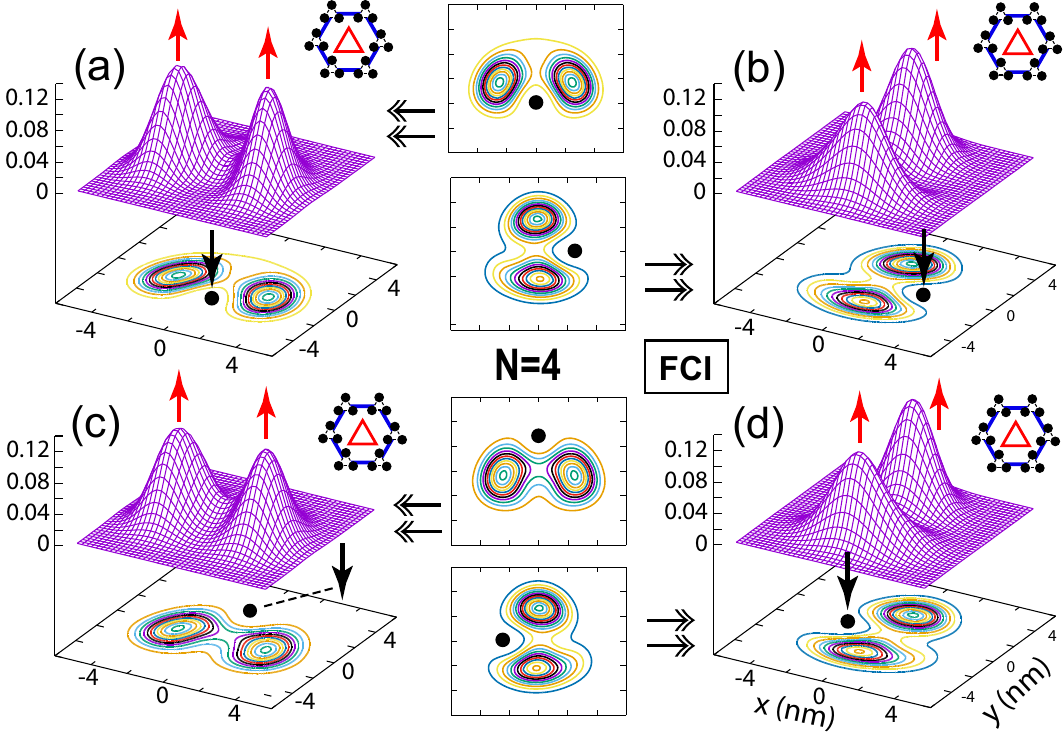}
\caption{
FCI spin-resolved CPDs ($P_{\uparrow\downarrow}$)  for the ground state (with total spin $S=1$ and 
total-spin projection $S_z=0$) for $N=4$ holes in the central MQD taking into 
consideration the crystal-field effect generated by the charge carriers in the surrounding six moir\'e 
pockets (see blue hexagon, with radius $a_M$, in the schematics). $\kappa=5$ in all four panels. 
 The black arrows indicate the chosen down spin of the fixed hole which is situated at the same
 positions $\br_0$ as in Fig.\ \ref{cpdn4dd}. The red arrows indicate that the plotted surfaces reveal 
 the distributions of a second spin-up hole. All other parameters are the same as those in Fig.\ 
 \ref{chdn4}(a), which portrays the corresponding charge density. The CPDs are in units of
 1/nm$^4$ and are normalized to 2/nm$^2$. 
}
\label{cpdn4du}
\end{figure*}

The spin-resolved CPD gives the spatial probability
 distribution of finding a second fermion with spin projection $\sigma$ under the condition that a first fermion is
 located (fixed) at the reference point  ${\bf r}_0$ with spin projection $\sigma_0$; $\sigma$ and $\sigma_0$ can
 be either up $(\uparrow$) or down ($\downarrow$). The meaning of a space-only CPD is analogous, but without
 consideration of spin. Mathematically, the spin-resolved CPDs are defined \cite{yann07,yann07.2} as
 \begin{equation}
P_{\sigma\sigma_0}({\bf r}, {\bf r}_0)=  \langle \Phi^{\rm{FCI}} |
\sum_{i \neq j} \delta({\bf r} - {\bf r}_i) \delta({\bf r}_0 - {\bf r}_j)
\delta_{\sigma \sigma_i} \delta_{\sigma_0 \sigma_j}
|\Phi^{\rm{FCI}}\rangle.
\label{sponcpd} 
\end{equation}

\begin{figure}[t]
\centering\includegraphics[width=8.1cm]{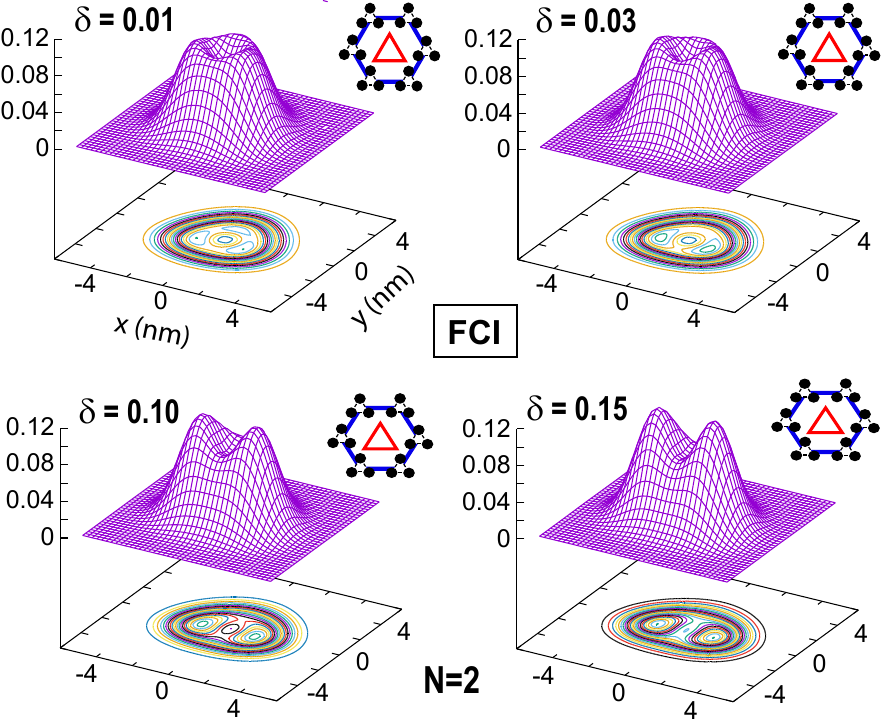}
\caption{
Evolution of the FCI charge densities for $N=2$ holes as a function of the strain parameter $\delta$,
whose value is displayed in each frame. The evolution from three to two peaks can be seen in the
contour plots in the $x-y$ plane, as well as in the 3D surfaces.
Remaining parameters: effective mass $m^*=0.90 m_e$, moir\'e
lattice constant $a_M=9.8$ nm, depth of a moir\'e potential pocket $v_0=10.3$ eV, $\phi=20^\circ$,
and $\kappa=5.0$. Charge densities are in units of 1/nm$^2$.
}
\label{strain}
\end{figure}

In Figs.\ \ref{cpdn4dd} and  \ref{cpdn4du}, we display SR-CPDs for the ground state of the 4-hole sliding WM 
whose charge density was displayed in Fig.\ \ref{chdn4}(a). Two different families of SR-CPDs are considered,
namely: (1) the $P_{\downarrow\downarrow}$ family, which can be described as "fix a hole with spin down and 
look for another hole with spin down, as well" (see Fig.\ \ref{cpdn4dd}), and (2) the $P_{\uparrow\downarrow}$ 
family, which can be described as "fix a hole with spin down and look for another hole with spin up" (see Fig.\
\ref{cpdn4du}). The remaining two SR-CPD families, $P_{\uparrow\uparrow}$ and $P_{\downarrow\uparrow}$,
can be simply obtained from those plotted in Figs.\ \ref{cpdn4dd} and \ref{cpdn4du} by flipping the spins.

These four families of SR-CPDs reveal an intrinsic structure for the sliding WM which consists of four localized holes, two with up spins and the other two with downs spins in an alternating configuration. 
Moreover, following similar reasoning as in Ref.\ \cite{yann09}, one can deduce that the intrinsic spin
eigenfunction associated with the ground-state sliding WM of Fig.\ \ref{chdn4}(a) is given by the
expression
\begin{align}
\chi^{S=1}_{S_z=0} = ( |\uparrow\downarrow\uparrow\downarrow\rangle -
 |\downarrow\uparrow\downarrow\uparrow\rangle)/\sqrt{2}.
\end{align}

Unlike the case of a circular confinement \cite{yann00,yann04,yann07}, the CPD surfaces
in  Figs.\ \ref{cpdn4dd} and \ref{cpdn4du} vary with the azimuthal angle.
As mentioned above, this is associated with the fact that the corresponding 
charge density in Fig.\ \ref{chdn4}(a) has a trilobally deformed doughnut shape.

For completeness, in Fig.\ \ref{cpdn2} (see Appendix \ref{a7}), we display SR-CPDs for the
singlet ground state of the 2-hole sliding WM whose charge density was displayed in Fig.\ \ref{chdn2ci}(a).

Finally, we comment briefly on the approaches that bridge the gap between the sS-UHF and FCI solutions.
In the case of a circular confinement, we showed earlier \cite{yann02.2,yann07} that this can be 
achieved by restoring the broken symmetry of the sS-UHF solutions via projection techniques.
This is equivalent to carrying out a mixing (superposition) of the equivalent (merely rotated) sS-UHF solutions
over all the azimuthal angles using coefficients supplied by the projection methodology. In the case
of a moir\'e quantum dot considered here, the mean-field components entering in the mixing are not  
equivalent in all the azimuthal directions, and thus the Griffin-Hill-Wheeler generator coordinate
method \cite{yann07,shei21,grif57,hill53}, which is broader in 
scope than the symmetry-restoration approach, must be employed \cite{yannun}.

\section{EXPLICIT CONSIDERATION OF STRAIN}
\label{a2}

In Fig.\ \ref{strain}, we display FCI CDs for $N=2$ holes as a function of the strain parameter
$\delta$. The remaining parameters are the same as for the unstrained ($\delta=0$) CD displayed in
Fig.\ \ref{chdn2ci}(a). The parameter $\delta$ is defined as in Ref.\ \cite{crom23}, namely, lengths along
the $x$-axis are extended by a factor $(1+\delta/2)$, whilst lengths along the $y$-axis are contracted
by a factor $(1-\delta/2)$.

\section{THE CONFIGURATION INTERACTION METHOD}
\label{a3}

The full configuration interaction (FCI) methodology has a long history, starting in quantum 
chemistry; see Refs.\ \cite{shav98,szabo}. The method was adapted to two dimensional problems and 
found extensive applications in the fields of semiconductor quantum dots
\cite{yann03,szaf03,ront06,yann07.2,yann07,yann09,yann22.2,yann22.3} and of the fractional quantum 
Hall effect \cite{yann04,yann21}.

Our 2D FCI is described in our earlier publications. The reader will find a comprehensive exposition
in Appendix B of Ref.\ \cite{yann22.2}, where the method was applied to GaAs double-quantum-dot 
quantum computer qubits. We specify that, in the application to moir\'{e} DQDs, we keep similar
space orbitals, $\varphi_j(x,y)$, $j=1,2,\ldots,K$, that are employed in the building of the 
single-particle basis of spin-orbitals used to construct the Slater determinants $\Psi_I$, which 
span the many-body Hilbert space [see Eq.\ (B4) in Ref.\ \cite{yann22.2}; the index $I$ counts the 
Slater determinants]. The orbitals $\varphi_j(x,y)$ are determined as solutions (in Cartesian coordinates) 
of the auxiliary Hamiltonian
\begin{align}
H_{\rm aux} = \frac{{\bf p}^2}{2 m^*} +  \frac{1}{2} m^* \omega_0^2 (x^2+y^2),
\label{haux}
\end{align}
where $m^*\omega_0^2=16 \pi^2 v_0 \cos(\phi)/a_M^2$, i.e., only the isotropic parabolic (harmonic)
contribution of the MQD confinement $V_{\rm MQD}(\br)$ [see Eqs.\ (2) and (4) in the main text] is included.

The matrix elements $\langle \varphi_i(x,y) | \sin(3\theta) r^3 | \varphi_j(x,y) \rangle$
of the anisotropic term in the MQD confinement are calculated analytically using the algebraic
language MATHEMATICA \cite{math22} and the Hermite-to-Laguerre (Cartesian-to-polar) transformations listed
in Ref.\ \cite{kime93}, whereas the matrix elements 
$\langle \varphi_i(x,y) | H_{\rm CF} | \varphi_j(x,y) \rangle$ of the crystal field are calculated numerically.

Following Ref.\ \cite{yann22.2}, we use a sparse-matrix eigensolver based on Implicitly Restarted 
Arnoldi methods to diagonalize the many-body Hamiltonian in Eq.\ (3) of the main text.
We stress that the FCI conserves both the total spin and the total-spin projection.


\begin{figure}[t]
\centering\includegraphics[width=8.0cm]{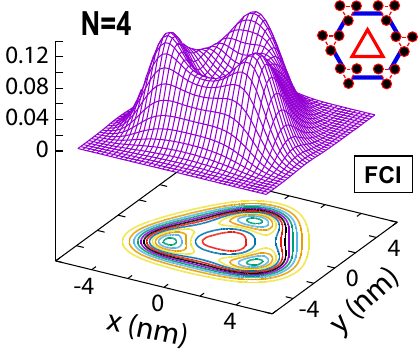}
\caption{
FCI charge density for the ground state of $N=4$ holes in the central 
MQD (represented by the red triangle in the schematic) taking into 
consideration the crystal-field effect generated by the charge carriers in the surrounding six moir\'e 
pockets (see blue hexagon, with radius $a_M$, in the schematic), but with an arrangement of the 
surrounding point charges  that is antiparallel to that of Fig.\ 3(a) in the main text
(see black dots in the schematic). $\kappa=5$.
All other parameters and notations are the same as those in Fig.\ 3(a) of the main text.
}
\label{S1}
\end{figure}

\begin{figure}[t]
\centering\includegraphics[width=8.0cm]{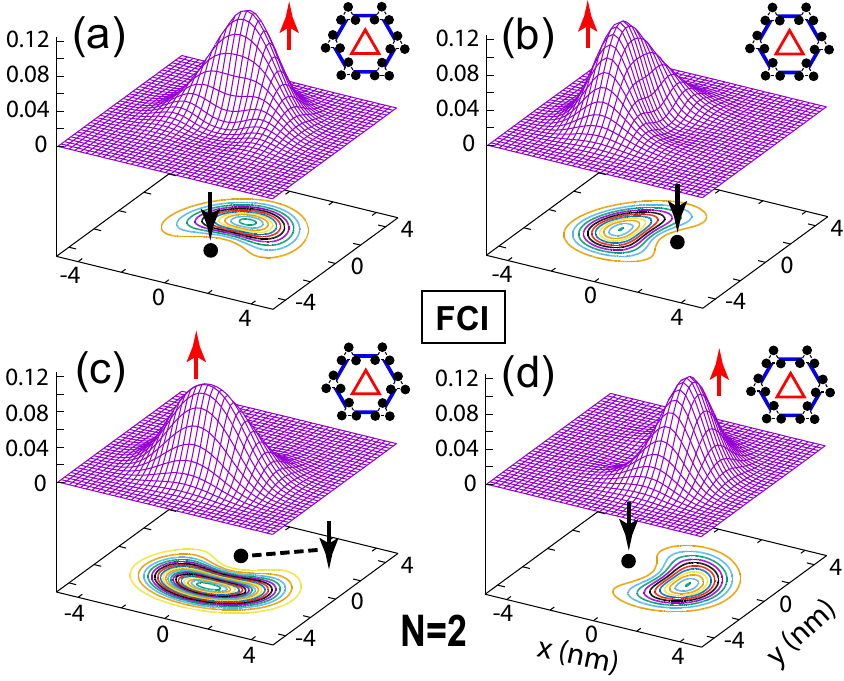}
\caption{
FCI spin-resolved CPDs ($P_{\uparrow\downarrow}$) for the singlet ground state (with total spin $S=0$
and total-spin projection $S_z=0$) for $N=2$ holes in the central MQD taking into
consideration the crystal-field effect generated by the charge carriers in the surrounding six moir\'e
pockets (see blue hexagon, with radius $a_M$, in the schematics). $\kappa=5$ in all four panels.
The black arrows indicate the chosen down spin of the fixed hole which is situated at the four
different positions $\br_0$ marked by a solid black dot: (a) $\br_0$=(0, -1.30 nm), (b) $\br_0$=(1.30 nm,
0), (c) $\br_0$=(0,1.30 nm), and $\br_0$=(-1.30 nm, 0). The red arrows indicate that the
plotted surfaces reveal the distributions of the second spin-up hole. All other details are the same as those
in Fig.\ 1(a), which portrays the corresponding charge density. The CPDs are in units of
1/nm$^4$ and are normalized to 1/nm$^2$.
}
\label{cpdn2}
\end{figure}

\section{THE SPIN-AND-SPACE UNRESTRICTED HARTREE-FOCK}
\label{a4}

Early on in the context of 2D materials, the spin-and-space unrestricted Hartree-Fock (sS-UHF) was 
employed in Ref.\ \cite{yann99} to describe formation of Wigner molecules at the mean-field level. 
This methodology employs the Pople-Nesbet equations \cite{szabo,yann07}. The sS-UHF WMs are
self-consistent solutions of the Pople-Nesbet equations that are obtained by relaxing both the 
total-spin and space symmetry requirements. For a detailed description of the Pople-Nesbet equations
in the context of three-dimensional natural atoms and molecules, see Ch.\ 3.8 in Ref.\ \cite{szabo}.
For a detailed description of the Pople-Nesbet equations in the context of two-dimensional
artificial atoms and semiconductor quantum dots, see Sec.\ 2.1 of Ref.\ \cite{yann07}.
We note that the Pople-Nesbet equations conserve the total-spin projection, but not the total spin.
Convergence of the self-consistent iterations was achieved in all cases by mixing the input and
output charge densities at each iteration step. The convergence criterion was set to a difference
of $10^{-12}$ meV between the input and output total UHF energies at the same iteration step.\\    
~~~~\\

\section{CHARGE DENSITIES FROM FCI AND UHF WAVE FUNCTIONS}
\label{a5}

The FCI single-particle density (charge density) is the expectation value of a one-body operator
\begin{equation}
\rho({\bf r}) = \langle \Phi^{\rm FCI}
\vert  \sum_{i=1}^N \delta({\bf r}-{\bf r}_i)
\vert \Phi^{\rm FCI} \rangle,
\label{elden}
\end{equation}
where $\Phi^{\rm FCI}$ denotes the many-body (multi-determinantal) FCI wave function, namely,
\begin{equation}
\Phi^{\rm FCI} ({\bf r}_1, \ldots , {\bf r}_N) =
\sum_I C_I \Psi_I({\bf r}_1, \ldots , {\bf r}_N),
\label{mbwf}
\end{equation}
with $\Psi_I({\bf r)}$ denoting the Slater determinants that span the many-body Hilbert space. 
The expansion coefficients $C_I$ are a byproduct of the exact diagonalization of the 
many-body Hamiltonian $H_{\rm MB}$.

For the sS-UHF case, one substitutes $\Phi^{\rm FCI}$ in Eq.\ (\ref{elden}) with the 
single-determinant, $\Psi^{\rm UHF}({\bf r})$, solution of the Pople-Nesbet equations. 
$\Psi^{\rm UHF}({\bf r})$ is built out from the UHF spin-orbitals whose space part has the form:
\begin{align}
u^\alpha_i=\sum_{\mu=1}^K {\cal C}^\alpha_{\mu i} \varphi_\mu, \;\;\; i=1,\ldots,K,
\end{align}
and
\begin{align}
u^\beta_i=\sum_{\mu=1}^K {\cal C}^\beta_{\mu i} \varphi_\mu, \;\;\; i=1,\ldots,K,
\end{align}
where the expansion coefficients ${\cal C}^\alpha_{\mu i}$ and ${\cal C}^\beta_{\mu i}$ are solutions
of the Pople-Nesbet equations, with $\alpha$ and $\beta$ denoting up and down spins, respectively.

The UHF expression for the charge density simplifies to:
\begin{align}
\rho^{\rm UHF}(\br)=\sum_{i=1}^{N^\alpha} |u^\alpha_i(\br)|^2 + \sum_{j=1}^{N^\beta} |u^\beta_j(\br)|^2,  
\end{align}
with ${N^\alpha}$, ${N^\beta}$ being the number of lowest-energy occupied spin-up and spin-down orbitals, respectively.\
~~~\\

\section{CASE OF AN ANTIPARALLEL ARRANGEMENT OF POINT CHARGES IN
THE SURROUNDING POCKETS}
\label{a6}

Fig.\ \ref{S1} displays the FCI charge density for the ground state of $N=4$ holes in the central 
MQD taking into account the crystal-field effect generated by the charge carriers in the surrounding 
six moir\'e pockets, but with an arrangement of the surrounding point charges that is antiparallel to 
that of Fig.\ 3(a) in the main text. The effect of such a flip is minimal, with the central charge density
retaining the same shape and orientation that is associated with formation of a sliding WM.
This behavior is connected to the fact that the overall crystal-field symmetry retains the
hexalobular $C_6$ character.

\section{CPDs FOR THE 2-HOLE SLIDING WIGNER MOLECULE}
\label{a7}

In Fig.\ \ref{cpdn2}, we display SR-CPDs for the singlet ground state of the 2-hole sliding WM whose charge
density was displayed in Fig.\ 1(a) of the main text. In all four panels of Fig.\ \ref{cpdn2}, the second hole
is found to follow in an antipodal arrangement the movement of the black dot that represnts the positions of
the fixed (reference) hole. We note that, unlike the case of a circular confinement \cite{yann00,yann04,yann07},
the CPD shapes can, and do indeed, vary with the azimuthal angle; e.g., compare the surfaces in Figs.\
\ref{cpdn2}(a) and \ref{cpdn2}(c). This variation is natural because the twodimensional integral of the CPD
(second-order correlation) over the fixed point must yield the deformed doughnut-like charge density
(first-order correlation) of Fig.\  1(a) \cite{lowd55}.

\nocite{*}
\bibliographystyle{apsrev4-2}
\bibliography{mycontrols,crystal_field_moire_arx2}

\end{document}